\documentclass[
    aps,
    pra,
    amsmath,amssymb,
    reprint,
    superscriptaddress
   ]{revtex4-1}
   \newcommand{\bv}[1]{
     \ensuremath{\boldsymbol{#1}}
   }
   \newcommand{\expect}[1]{
     \ensuremath{\langle{#1}\rangle}
   }
   \newcommand{\Tr}{\text{Tr}}
   
   \usepackage{graphicx}% Include figure files
   \usepackage{color}
   \usepackage{dcolumn}% Align table columns on decimal point
   \usepackage{siunitx}
   \usepackage{bm}% bold math
   \usepackage{braket}
   \begin{document}
   
   %\preprint{AIP/123-QED}
   
   \title[]{Quantum Circuit Learning}
   
   \author{K. Mitarai}
   \email{mitarai@qc.ee.es.osaka-u.ac.jp}
   \affiliation{Graduate School of Engineering Science, Osaka University, 1-3 Machikaneyama, Toyonaka, Osaka 560-8531, Japan.}
   \author{M. Negoro}
   \affiliation{Graduate School of Engineering Science, Osaka University, 1-3 Machikaneyama, Toyonaka, Osaka 560-8531, Japan.}
   \affiliation{JST, PRESTO, 4-1-8 Honcho, Kawaguchi, Saitama 332-0012, Japan}
   \author{M. Kitagawa}
   \affiliation{Graduate School of Engineering Science, Osaka University, 1-3 Machikaneyama, Toyonaka, Osaka 560-8531, Japan.}
   \affiliation{Quantum Information and Quantum Biology Division, Institute for Open and Transdisciplinary Research Initiatives, Osaka University}  
   \author{K. Fujii}
   \email{fujii.keisuke.2s@kyoto-u.ac.jp}
   \affiliation{Graduate School of Science, Kyoto University, Yoshida-Ushinomiya-cho, Sakyo-ku, Kyoto 606-8302, Japan.}
   \affiliation{JST, PRESTO, 4-1-8 Honcho, Kawaguchi, Saitama 332-0012, Japan}

   \date{\today}
   
   \begin{abstract}
       We propose a classical-quantum hybrid algorithm for machine learning on near-term quantum processors, which we call quantum circuit learning.
       A quantum circuit driven by our framework learns a given task by tuning parameters implemented on it.
       The iterative optimization of the parameters allows us to circumvent the high-depth circuit.
       Theoretical investigation shows that a quantum circuit can approximate nonlinear functions, which is further confirmed by numerical simulations.
       Hybridizing a low-depth quantum circuit and a classical computer for machine learning, the proposed framework paves the way toward applications of near-term quantum devices for quantum machine learning.
   \end{abstract}
   
   \pacs{Valid PACS appear here}
   \maketitle
   
    \section{Introduction}
    In recent years, machine learning has acquired much attention in a wide range of areas including the field of quantum physics \cite{Carleo2017, Rupp2011, Broecker2016, Ramakrishnan2015, August2016}.
    Since quantum information processing is expected to bring us exponential speedups on some problems \cite{Shor1997, Nielsen2010},
    usual machine learning tasks might as well be improved when it is carried on a quantum computer.
    Also, for the purpose of learning a complex quantum system, it is natural to utilize a quantum system as our computational resource.
    A variety of machine learning algorithms for quantum computers has been proposed \cite{Kerenidis2017, Wiebe2012, Rebentrost2014, Cao2017}, since Harrow-Hassidim-Lloyd (HHL) algorithm \cite{Harrow2009} enabled us to perform basic matrix operations on a quantum computer.
    These HHL-based algorithms have the quantum phase estimation algorithm \cite{Nielsen2010} at its heart, which requires a high-depth quantum circuit.
    To circumvent a high-depth quantum circuit, which is still a long-term goal on the hardware side,
    classical-quantum hybrid algorithms consisting of a relatively low-depth quantum circuit such as quantum variational eigensolver \cite{Peruzzo2013, Kandala2017} (QVE) and
    quantum approximate optimization algorithm \cite{Farhi2014,Farhi2016,Otterbach2017} (QAOA) have been suggested.
    In these methods, a problem is encoded into a Hermitian matrix \(A\).
    Its expectation value \(\expect{A}\) with respect to an ansatz state \(\ket{\psi(\bv{\theta})}\) is iteratively optimized by tuning the parameter \(\bv{\theta}\).
    The central idea of hybrid algorithms is dividing the problem into two parts, each of which can be performed easily on a classical and a quantum computer.
    
    In this paper, we present a new hybrid framework, which we call quantum circuit learning (QCL), for machine learning with a low-depth quantum circuit.
    In QCL, we provide input data to a quantum circuit, and iteratively tunes the circuit parameters so that it gives the desired output.
    Gradient-Based systematic optimization of parameters is introduced for the tuning just like backpropagation method \cite{Bishop2013} utilized in feedforward neural networks.
    We theoretically show that a quantum circuit driven by the QCL framework can approximate any analytical function if the circuit has a sufficient number of qubits.
    The ability of the QCL framework to learn nonlinear functions and perform a simple classification task is demonstrated by numerical simulations.
    Also, we show by simulation that a 6-qubit circuit is capable of fitting dynamics of 3 spins out of a 10-spin system with fully connected Ising Hamiltonian.
    We stress here that the proposed framework is easily realizable on near-term devices.
       
    \section{Quantum circuit learning}
    \subsection{Algorithm}
    Our QCL framework aims to perform supervised or unsupervised learning tasks \cite{Bishop2013}.
    In supervised learning, an algorithm is provided with a set of input \(\{\bv{x}_i\}\) and corresponding teacher data \(\{f(\bv{x}_i)\}\).
    The algorithm learns to output \(y_i = y(\bv{x}_i,\bv{\theta})\) that is close to the teacher  \(f(\bv{x}_i)\), by tuning \(\bv{\theta}\).
    The output and the teacher can be vector-valued.
    QCL assigns the calculation of the output \(y_i\) to a quantum circuit and the update of the parameter \(\bv{\theta}\) to a classical computer.
    The objective of learning is to minimize a cost function, which is a measure of how close the teacher and the output is, by tuning \(\bv{\theta}\).
    As an example, the quadratic cost \(L = \sum_i \|f(\bv{x}_i) - y_i\|^2\) is often used in regression problems.
    On the other hand, in unsupervised learning (e.g. clustering), only input data are provided, and some objective cost function that does not involve teacher is minimized.

    Here we summarize the QCL algorithm on \(N\) qubit circuit:
    \begin{enumerate}
        \item Encode input data \(\{\bv{x}_i\}\) into some quantum state \(\ket{\psi_{\text{in}}(\bv{x}_i)}\) by applying a unitary input gate \(U(\bv{x}_i)\) to initialized qubits \(\ket{0}\)
        \item Apply a \(\bv{\theta}\)-parameterized unitary \(U(\bv{\theta})\) to the input state and generate an output state \(
            \ket{\psi_{\text{out}}(\bv{x}_i,\bv{\theta})} = U(\bv{\theta})\ket{\psi_\text{in}(\bv{x}_i)}\).
        \item Measure the expectation values of some chosen observables. Specifically, we use a subset of Pauli operators \(\{B_j\} \subset \{I,X,Y,Z\}^{\otimes N}\). Using some output function \(F\), output \(y_i = y(\bv{x}_i,\bv{\theta})\) is defined to be
        \(
            y(\bv{x}_i,\bv{\theta}) \equiv F\left(\{\expect{B_j(\bv{x}_i,\bv{\theta})}\}\right).
        \)
        \item Minimize the cost function \(L\left(f(\bv{x}_i),y(\bv{x}_i,\bv{\theta})\right)\) of the teacher \(f(\bv{x}_i)\) and the output \(y_i\), by tuning the circuit parameters \(\bv{\theta}\) iteratively.
        \item Evaluate the performance by checking the cost function with respect to a data set that is taken independently from the training one.
    \end{enumerate}
    
    \subsection{Relation with existing algortihms}
    Minimization of the quadratic cost can be performed using a high-depth quantum circuit with HHL-based algorithms. For example, Ref. \cite{Schuld2016} shows a detailed procedure.
    This matrix inversion approach is similar to the quantum support vector machine \cite{Rebentrost2014}.
    As opposed to this, QCL applied to a regression problem minimizes the cost by iterative optimization, successfully circumventing a high-depth circuit.

\begin{figure}
        \includegraphics[width=0.9\linewidth]{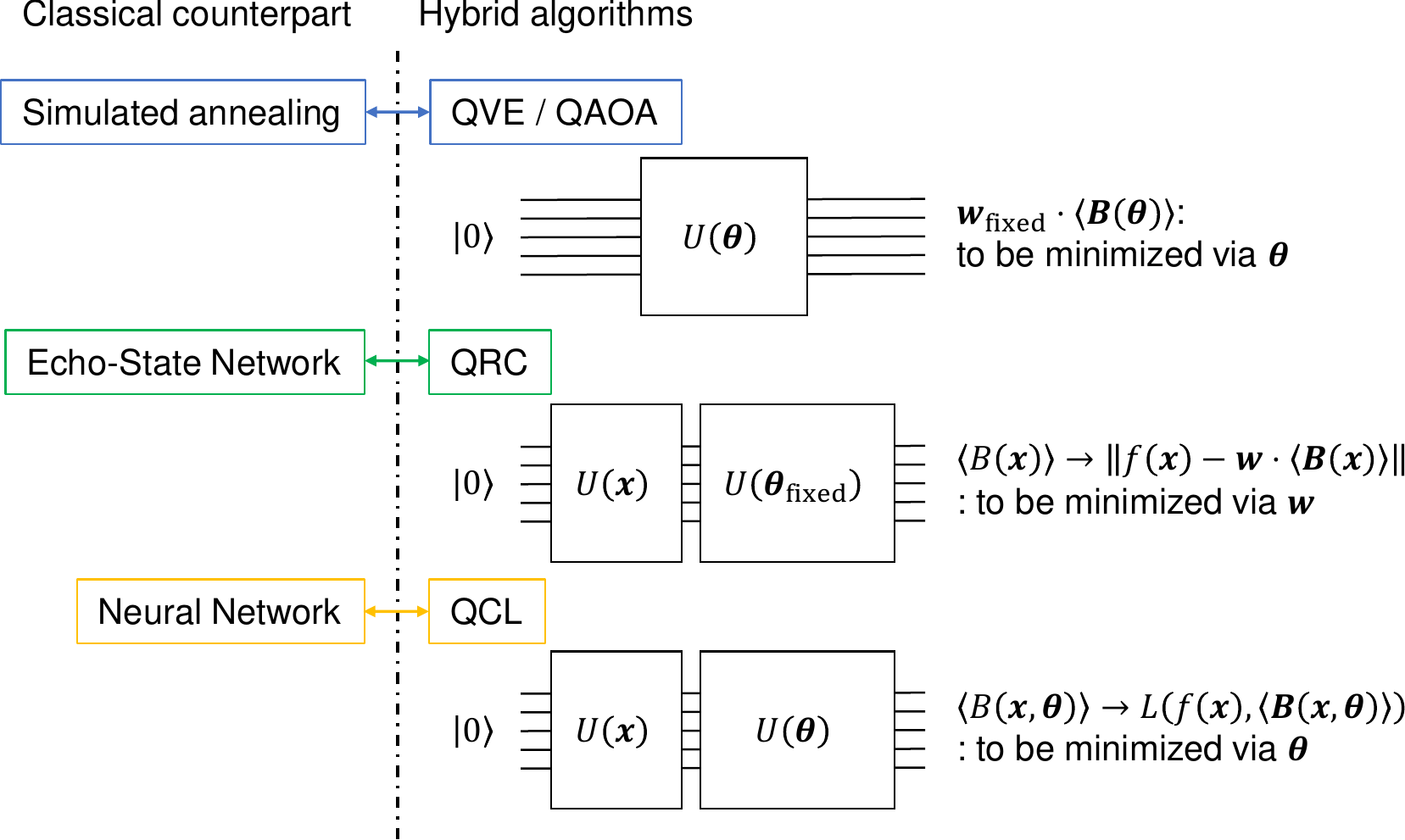}
        \caption{\label{comp} Comparison of QVE/QAOA, QRC, and presented QCL framework.
        In QVE, the output of the quantum circuit is directly minimized.
        QRC and QCL both optimize the output to the teacher \(f(\bv{x})\).
        QRC optimization is done via tuning    the linear weight \(\bv{w}\), as opposed to QCL approach which tunes the circuit parameter \(\bv{\theta}\).}
    \end{figure}

    Quantum reservoir computing (QRC) \cite{Fujii2016} shares a similar idea, in a sense that it passes the central optimization procedure to a classical computer.
    There, output is defined to be \(y(\bv{x}_i) \equiv \bv{w}\cdot\expect{\bv{B}}\) where \(\bv{B}\) is a set of observables taken from quantum many-body dynamics driven with a fixed Hamiltonian,
    and \(\bv{w}\) is the weight vector, which is tuned on a classical device to minimize a cost function.
    The idea stems from a so-called echo-state network approach \cite{Jaeger2004}.
    If one views QRC as a quantum version of the echo-state network, QCL, which tunes the whole network, can be regarded as a quantum counterpart of a basic neural network.
    In QVE/QAOA, the famous hybrid quantum algorithms, weighted sum of measured expectation values \(\bv{w}_{\text{fixed}}\cdot\expect{\bv{B}(\bv{\theta})}\) is minimized by tuning the parameter \(\bv{\theta}\).
    There, an input \(\bv{x}\) of a problem, such as geometry of a molecule or topology of a graph, is encoded to the weight vector \(\bv{w}_{\text{fixed}}\) as \(\bv{w}_{\text{fixed}}(\bv{x})\).
    This procedure corresponds to a special case of QCL where we do not use the input unitary \(U(\bv{x})\), and a cost function \(L = \bv{w}_{\text{fixed}}\cdot\expect{\bv{B}}\) is utilized.
    Fig. \ref{comp} summarizes and shows the comparison of QVE/QAOA, QRC, and presented QCL framework.

    \subsection{Ability to approximate a function}
    First, we consider the case where input data are one dimension for simplicity. It is straightforward to generalize the following argument for higher dimensional inputs.

    Let \(x\) and \(\rho_{\text{in}}(x) = \ket{\psi_{\text{in}}(x)}\bra{\psi_{\text{in}}(x)}\) be an input data and a corresponding density operator of input state.
    \(\rho_{\text{in}}(x)\) can be expanded by a set of Pauli operators \(\{P_k\} = \{I,X,Y,Z\}^{\otimes N}\) with \(a_k(x)\) as coefficients,
    \(\rho_{\text{in}}(x) = \sum_k a_k(x)P_k\).
    A parameterized unitary transformation \(U(\bv{\theta})\) acting on \(\rho_{\text{in}}(x)\) creates the output state, which can also be expanded by \(\{P_k\}\) with \(\{b_k(x,\bv{\theta})\}\).
    Now let \(u_{ij}(\bv{\theta})\) be such that \(b_m(x,\bv{\theta}) = \sum_k u_{mk}(\bv{\theta})a_k(x)\).
    \(b_m\) is an expectation value of a Pauli observable itself, therefore, the output is linear combination of input coefficient functions \(a_k\) under unitarity constraints imposed on \(\{u_{ij}\}\).
    
    When the teacher \(f(x)\) is an analytical function, we can show, at least in principle, QCL is able to approximate it by considering a simple case with an input state created by single-qubit rotations.
    The tensor product structure of quantum system plays an important role in this analysis.
    Let us consider a state of \(N\) qubits:
    \begin{equation}\label{1qubitstate}
        \rho_{\text{in}}(x) = \frac{1}{2^N}\bigotimes_{i=1}^N\left[I + xX_i + \sqrt{1-x^2}Z_i\right].
    \end{equation}
    This state can be generated for any \(x\in [-1,1]\) with single-qubit rotations, namely, \(\prod_{i=1}^N R_{i}^Y (\sin^{-1}x),\) where \(R_{i}^Y(\phi)\) is the rotation of \(i\)th qubit around \(y\) axis with angle \(\phi\).
    The state given by Eq.~(\ref{1qubitstate}) has higher order terms up to the \(N\)th with respect to \(x\).
    Thus an arbitrary unitary transformation on this state can provide us with an arbitrary \(N\)th order polynomial as expectation values of an observable.
    Terms like \(x\sqrt{1-x^2}\) in Eq.~(\ref{1qubitstate}) can enhance its ability to approximate a function.
    
    Important notice in the example given above is that the highest order term \(x^N\) is hidden in an observable \(X^{\otimes N}\).
    To extract \(x^N\) from Eq.~(\ref{1qubitstate}), one needs to transfer the nonlocal observable \(X^{\otimes N}\) to a single-qubit observable using entangling gate such as the controlled-NOT gate.
    Entangling nonlocal operations are the key ingredients of the nonlinearity of an output.
    
    The above argument can readily be generalized to multi-dimensional inputs.
    Assume that we are given with \(d\)-dimensional data \(\bv{x} = \{x_1,x_2,..,x_d\}\) and want higher terms up to the \(n_k\)th \((k=1,\cdots,d)\) for each data,
    then encode this data into a \(N=\sum_k n_k\)-qubit quantum state as \(\rho_{\text{in}}(\bv{x}) = \frac{1}{2^N}\bigotimes_{k=1}^d\left(\bigotimes_{i=1}^{n_k}\left[I + x_kX_{i} + \sqrt{1-x_k^2}Z_{i}\right]\right).\)
    These input states automatically has an exponentially large number of independent functions as coefficient set to the number of qubits.
    The tensor product structure of quantum system readily ``calculates" the product such as \(x_1x_2\). 

    The unitarity condition of \(u_{ij}\) may have an effect to avoid an overfitting problem, which is crucial for their performance in machine learning or in regression methods. 
    One way to handle it in classical machine learning methods is adding a regularization term to the cost function.
    For example, ridge regression adds regularization term \(\|\bv{w}\|^2\) to the quadratic cost function.
    Overall \(L = \sum_i \|f(\bv{x}_i)-\bv{w}\cdot\bv{\phi}(\bv{x}_i)\|^2 + \|\bv{w}\|^2\) is minimized.
    The weight vector \(\bv{w}\) corresponds to the matrix element \(u_{ij}\) in QCL.
    The norm of a row vector \(\|\bv{u}_{i}\|\), however, is restricted to unity by the unitarity condition, which prevents overfitting, from the unitarity of quantum dynamics.    
    Simple examples of this are given in the Appendix.

    \subsection{Possible quantum advantages}
    We have shown by above discussions that approximation of any analytical functions is possible with the use of nonlinearity created by the tensor product.
    In fact, nonlinear basis functions are crucial for many methods utilized in classical machine learning.
    They require a large number of basis functions to create a complex model that predicts with high precision.
    However, the computational cost of learning increases with respect to the increasing number of basis functions.
    To avoid this problem, the so-called kernel trick method, which circumvents the direct use of a large number of them, is utilized \cite{Bishop2013}.
    In contrast, QCL directly utilizes the exponential number of functions with respect to the number of qubits to model the teacher, which is basically intractable on classical computers.
    This is a possible quantum advantage of our framework, which was not obvious from the previous approaches like QVE or QAOA.
    
    Moreover, let us now argue about the potential power of QCL representing complex functions.
    Suppose we want to learn the output of QCL that is allowed to use an unlimited resource in the learning process, via classical neural networks.
    Then it has to learn the relation between inputs and outputs of a quantum circuit, which, in general, includes universal quantum cellular automata \cite{Raussendorf2004, Janzing2005}.
    This certainly could not be achieved using a polynomial-size classical computational resource to the size (qubits and gates) of QCL.
    This implies that QCL has a potential power to represent more complex functions than the classical counterpart.
    Further investigations are needed including the learning costs and which actual learning problem enjoys such an advantage.

    \subsection{Optimization procedure}
    In QVE \cite{Peruzzo2013}, it has been suggested to use gradient-free methods like Nelder-Mead.
    However, gradient-based methods are generally more preferred when the parameter space becomes large.
    In neural networks, backpropagation method \cite{Bishop2013}, which is basically gradient descent, is utilized in the learning procedure.
    
    To calculate a gradient of an expectation value of an observable with respect to a circuit parameter \(\bv{\theta}\),
    suppose the unitary \(U(\bv{\theta})\) consists of a chain of unitary transformations \(\prod_{j=1}^l U_j(\theta_j)\) on a state \(\rho_{\text{in}}\) and we measure an observable \(B\).
    For convenience, we use notation \(U_{j:k} = U_j\cdots U_k\). Then \(\expect{B(\bv{\theta})}\) is given as \(\expect{B(\bv{\theta})} = \Tr\left(BU_{l:1}\rho_{\text{in}} U_{l:1}^\dagger \right).\)
    We assume \(U_j\) is generated by a Pauli product \(P_j\), that is, \(U_j(\theta) = \exp(-i\theta_jP_j/2)\).
    The gradient is calculated to be
    \(\frac{\partial\expect{B}}{\partial\theta_j} = -\frac{i}{2}\Tr\left(B U_{l:j}[P_j, U_{j-1:1}\rho_{\text{in}} U_{j-1:1}^\dagger]U_{l:j}^\dagger\right).\)
    While we cannot evaluate the commutator directly, the following property of commutator for an arbitrary operator \(\rho\) enables us to compute the gradient on a quantum circuit:
    \begin{equation}
        [P_j, \rho] = i\left[U_j\left(\frac{\pi}{2}\right)\rho U_j^\dagger\left(\frac{\pi}{2}\right) - U_j\left(-\frac{\pi}{2}\right)\rho U_j^\dagger\left(-\frac{\pi}{2}\right)\right].
    \end{equation}
    The gradient can be evaluated by
    \begin{align}
        \frac{\partial\expect{B}}{\partial\theta_j}&=
        \frac{1}{2}\Tr\left[BU_{l:j+1}U_j\left(\frac{\pi}{2}\right) \rho_{j} U_j^\dagger \left(\frac{\pi}{2}\right) U_{l:j+1}^\dagger\right]\nonumber\\
        \label{expect_gradient}&\quad-\frac{1}{2}\Tr\left[BU_{l:j+1}U_j\left(-\frac{\pi}{2}\right) \rho_{j} U_j^\dagger \left(-\frac{\pi}{2}\right) U_{l:j+1}^\dagger\right],
    \end{align}
    where \(\rho_j = U_{j:1}\rho_{\text{in}}U_{j:1}^\dagger\).
    Just by inserting \(\pm\pi/2\) rotation generated by \(P_j\) and measuring the respective expectation values \(\expect{B}^{\pm}_j\),
    we can evaluate the exact gradient of an observable \(\expect{B}\), via \(\frac{\partial\expect{B}}{\partial\theta_j} = \frac{\expect{B}^+_j - \expect{B}^-_j}{2}.\)
    A similar method is used by Li {\it et al.} \cite{Li2017} in their research of control pulse optimization with target quantum system.

    \section{Numerical simulations}
\begin{figure}
        \includegraphics[width=0.65\linewidth]{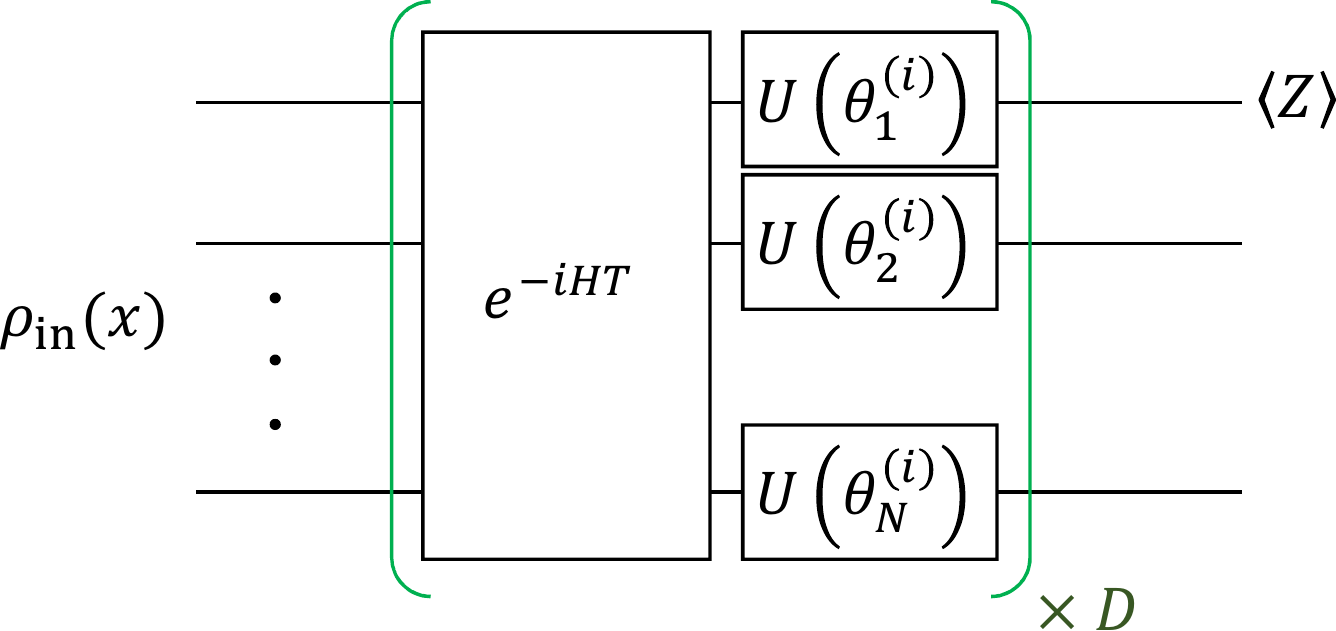}
        \caption{\label{simulated_circuit} Quantum circuit used in numerical simulations. The parameter \(\bv{\theta}\) of single qubit arbitrary unitaries \(U(\theta_j^{(i)})\) are optimized to minimize the cost function. \(D\) denotes the depth of the cicuit.}
    \end{figure}

    We demonstrate the performance of QCL framework for several prototypical machine learning tasks by numerically simulating a quantum circuit in the form of Fig.~\ref{simulated_circuit} with \(N=6\) and \(D=6\).
    \(U(\theta_j^{(i)})\) in Fig.~\ref{simulated_circuit} is an arbitrary rotation of a single qubit. We use the decomposition \(U(\theta_j^{(i)}) = R_j^X(\theta_{j1}^{(i)})R_j^Z(\theta_{j2}^{(i)})R_j^X(\theta_{j3}^{(i)})\).
    \(H\) is Hamiltonian of a fully connected transverse Ising model:
    \begin{equation}\label{Hamiltonian}
        H = \sum_{j=1}^N a_j X_j + \sum_{j=1}^N\sum_{k=1}^{j-1} J_{jk} Z_jZ_k.
    \end{equation}
    The coefficients \(a_j\) and \(J_{jk}\) are taken randomly from uniform distribution on \([-1,1]\). Evolution time \(T\) is fixed to 10.
    The results shown throughout this section are generated by the Hamiltonian with the same coefficients.
    Here we note that we have checked a similar result can be achieved with different Hamiltonians.
    The dynamics under this form of Hamiltonian can generate a highly entangled state and is, in general for a large number of qubits, not efficiently simulatable on a classical computer.
    Eq.~(\ref{Hamiltonian}) is the basic form of interaction in trapped ions or superconducting qubits, which makes the time evolution easily implementable experimentally.
    \(\bv{\theta}\) is initialized with random numbers uniformly distributed on \([0,2\pi]\).
    In all numerical simulations, outputs are taken from \(Z\) expectation values.
    To emulate a sampling, we added small gaussian noise with standard deviation \(\sigma\) determined by \(\sigma = \sqrt{2/N_s}(\expect{Z}^2-1)/4 \), where \(N_s\) and \(\expect{Z}\) are the number of samples and a calculated expectation value, to \(\expect{Z}\).
    \footnote{
        The simulation is carried on using Python library QuTip \cite{Johansson2013}.
        We use BFGS method \cite{Nocedal2006} provided in SciPy optimization library for optimization of parameters.
    }
    
    First, we perform fitting of \(f(x) = x^2, e^x, \sin x, |x|\) as a demonstration of representability of nonlinear functions \cite{Bishop2013}.
    We use the normal quadratic loss for the cost function.
    The number of teacher samples is 100.
    The output is taken from \(Z\) expectation value of the first qubit as shown in Fig. \ref{simulated_circuit}.
    In this simulation, we allow output to be multiplied by a constant \(a\) which is initialized to unity.
    This constant \(a\) and \(\bv{\theta}\) are simultaneously optimized.
    Input state \( \rho_{\text{in}}(x) \) is prepared by applying \(U_{\text{in}}(x) = \prod_j R_{j}^Z(\cos^{-1}x^2)R_{j}^Y(\sin^{-1}x)\) to initialized qubits \(\ket{0}\).
    This unitary creates a state similar to Eq.~(\ref{1qubitstate}).
    
\begin{figure}
        \includegraphics[width=\linewidth]{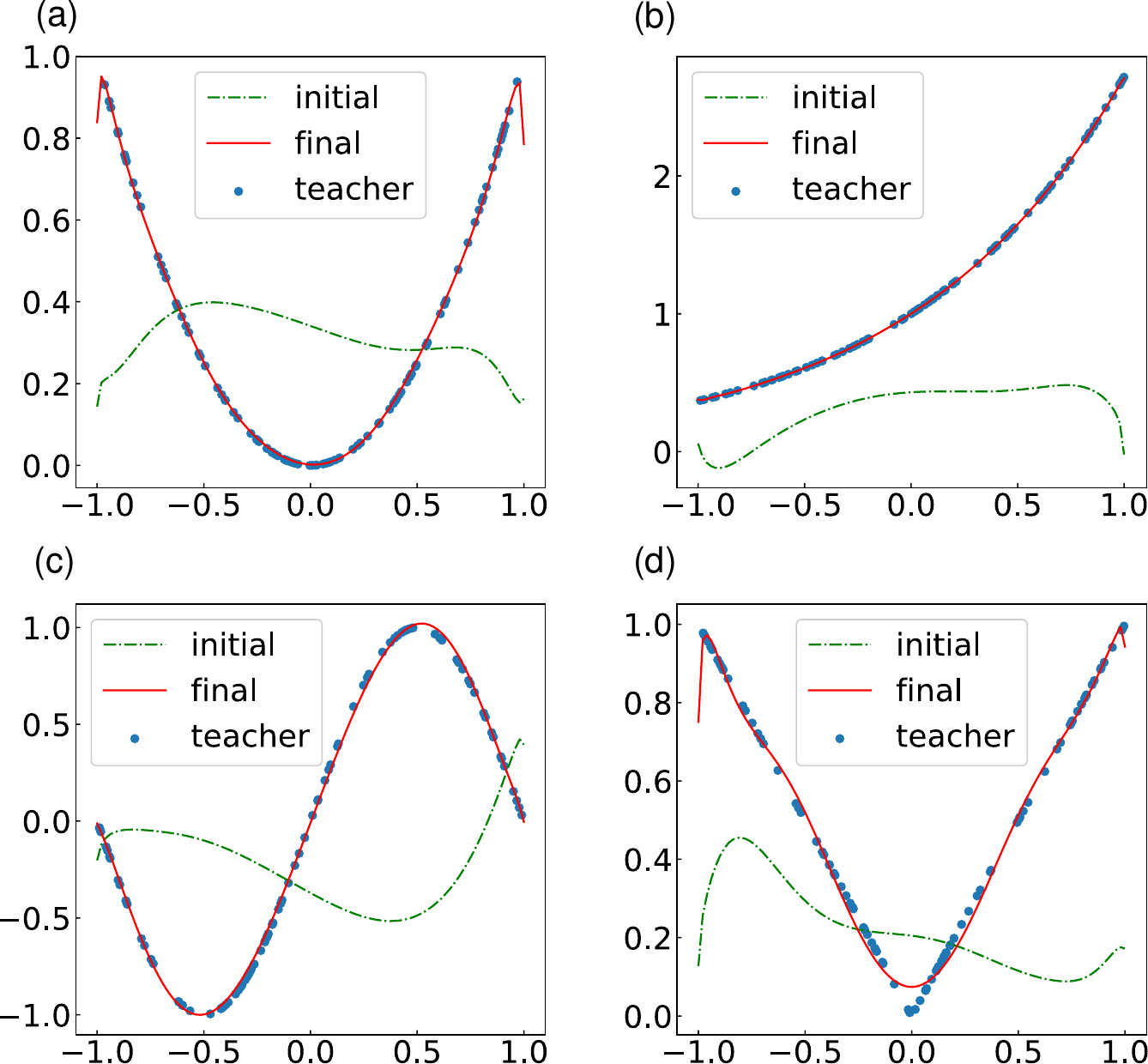}
        \caption{\label{fit} Demonstration of QCL performance to represent functions. ``initial" shows the output of quantum circuit with randomly chosen \(\theta\), and ``final" is the output from optimized quantum circuit. Each graph shows fitting of (a) \(x^2\), (b) \(e^x\), (c) \(\sin x\), (d) \(|x|\).}
    \end{figure}

    Results are shown in Fig. \ref{fit}.
    All of the functions are well approximated by a quantum circuit driven by presented QCL framework. 
    To approximate highly nonlinear functions such as \(\sin x\) or a nonanalytical function \(|x|\), QCL has brought out the high order terms which are initially hidden in nonlocal operators.
    The result of fitting \(|x|\) (Fig. \ref{fit} (d)) is relatively poor because of its nonanalytical characteristics.
    A possible solution for this is to employ different functions as an input function, such as Legendre polynomials.
    Although the choice of input functions affects the performance of QCL, the result shows that QCL with simple input has an ability to output a wide variety of functions.

\begin{figure}
    \includegraphics[width=\linewidth]{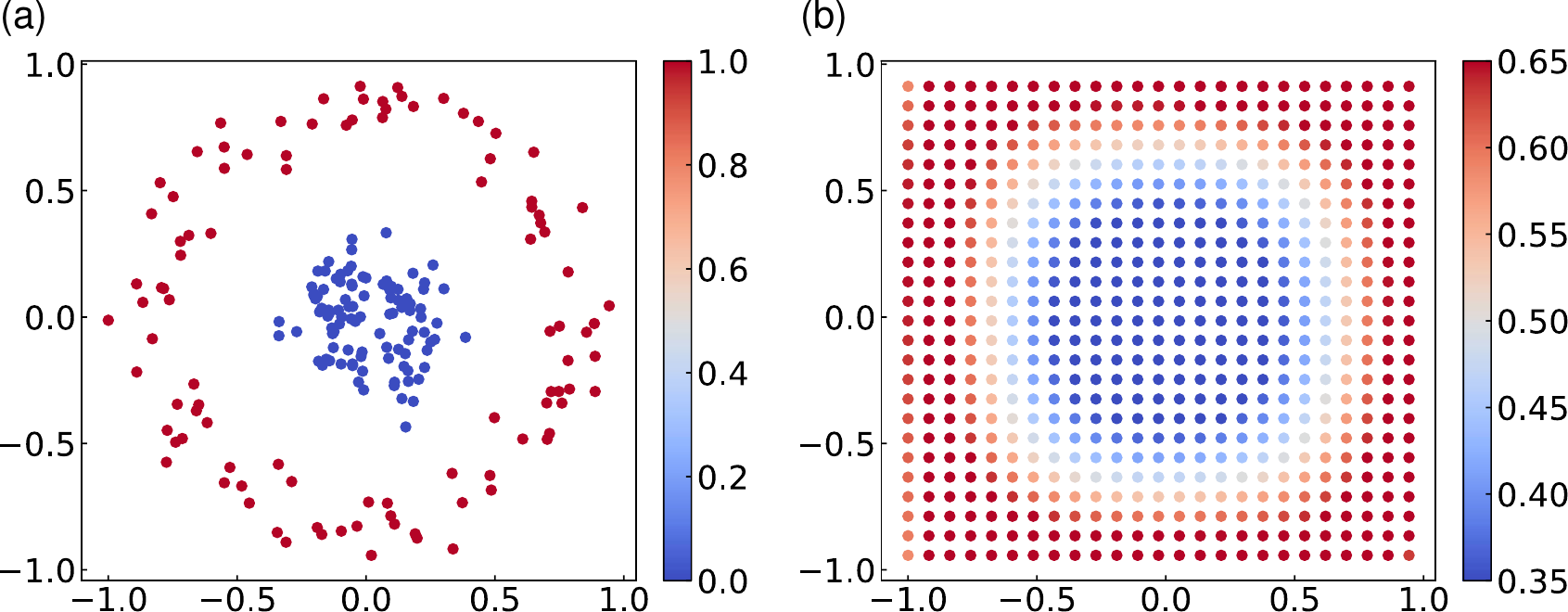}
    \caption{\label{circle_class}  Demonstration of a simple nonlinear classification task.
    (a) teacher data. Data points that belong to class 0, 1 is shown as blue and red dot, respectively.
    (b) Optimized output from first qubit (after softmax transformation).
    0.5 is the threshold for classification, less than and greater than 0.5 means that the point is classified as class 0 and 1, respectively.}
    \end{figure}

    As a second demonstration, the classification problem, which is an important family of tasks in machine learning, is performed.
    Fig. \ref{circle_class} (a) shows the training data set, blue and red points indicate class 0 and 1 respectively. Here we train the quantum circuit to classify based on each training input data points \(\bv{x}_i = (x_{i,0},x_{i,1})\).
    We define the teacher \(f(\bv{x}_i)\) for each input \(\bv{x}_i\) to be two dimensional vector \((1,0)\) for class 0, and \((0,1)\) for class 1.
    The number of teacher samples is 200 (100 for class 0, and 100 for class 1).
    The output is taken from the expectation value of the Pauli \(Z\) operator of the first 2 qubits, and they are transformed by softmax function \(\bv{F}\).
    For \(d\)-dimensional vector \(\bv{q}\), softmax function returns \(d\)-dimensional vector \(\bv{F}(\bv{q})\) with its \(k\)th element being \(F_k(\bv{q}) = e^{q_k}/\sum_i e^{q_i}\).
    Thus the output \(\bv{y}_i = (y_{i,0},y_{i,1})\) is defined by \(\bv{y}_i = \bv{F}(\expect{Z_1(\bv{x}_i,\bv{\theta}}),\expect{Z_2(\bv{x}_i,\bv{\theta}}))\) 
    For the cost function, we use the cross-entropy \(L = \sum_i \sum_{k\in\{0,1\}} \left(f(\bv{x}_{i})\right)_k\log y_{ik}\).
    The input state is prepared by applying \(U_{\text{in}}(x) = \prod_j R_{j}^Z(\cos^{-1}x_{i,j \text{ mod } 2}^2)R_{j}^Y(\sin^{-1}x_{i,j \text{ mod }2})\) to initialized qubits \(\ket{0}\). \(j \text{ mod } 2\) is the remainder of \(j\) devided by 2.
    In this task, the multiplication constant \(a\) is fixed to unity.
    
    Learned output is shown in Fig. \ref{circle_class} (b).
    We see that QCL works as well for the nonlinear classification task.
    The same task can be classically performed using, for example, kernel-trick support vector machine.
    Kernel-trick approach discards the direct use of a large number of basis functions with respect to the number of qubits, as opposed to QCL approach, which utilizes an exponentially large number of basis functions under certain constraints.
    In this sense, QCL can benefit from the use of a quantum computer.

    \begin{figure}
        \includegraphics[width=\linewidth]{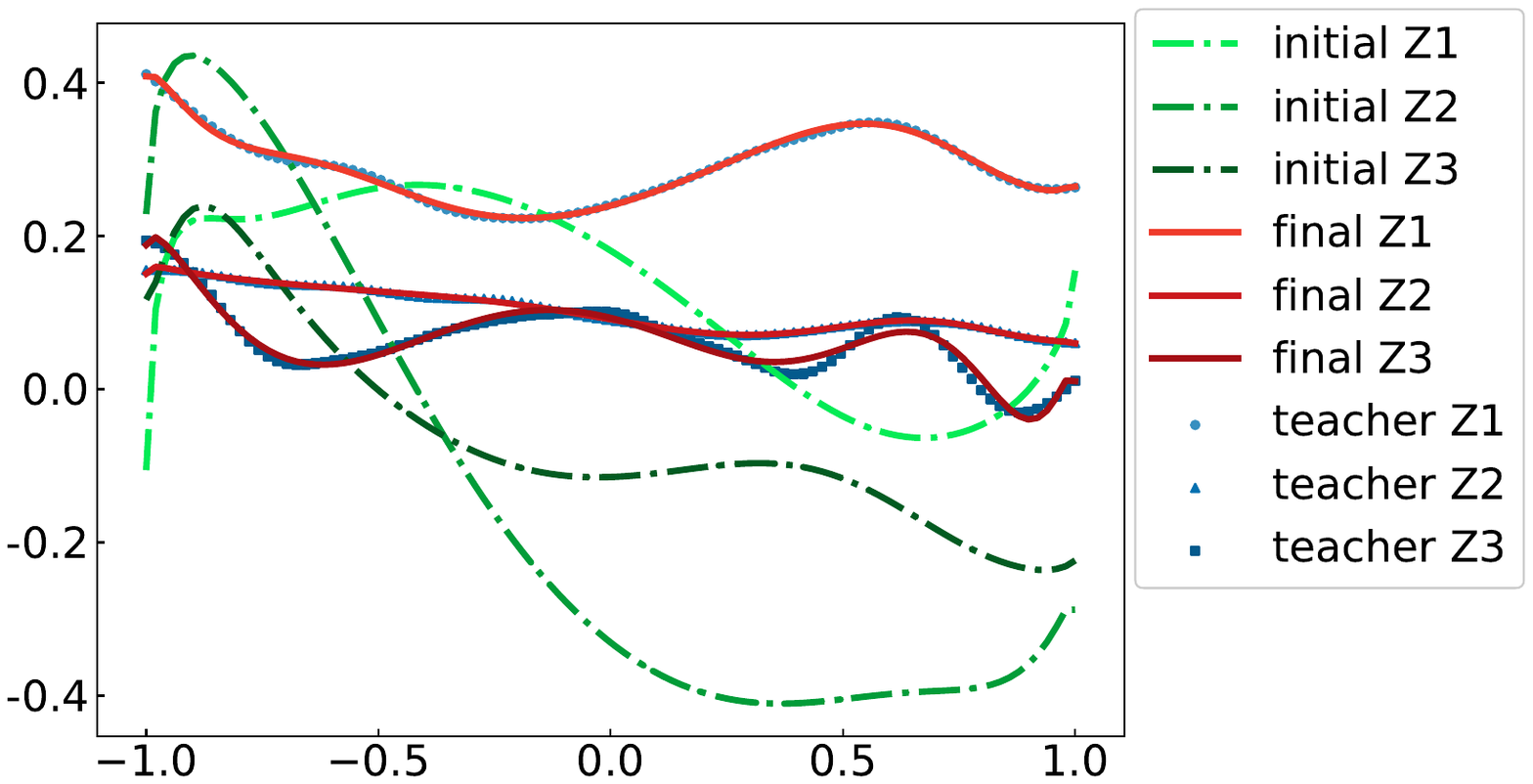}
        \caption{\label{dynamics_fit} Demonstration of fitting quantum many-body dynamics.
        Partial dynamics of a 10-spin system can be well approximated by a 6-qubit circuit.}
    \end{figure}

    Finally, we demonstrate the ability of QCL to perform a fitting task of quantum many-body dynamics.
    Simulation of dynamics of the 10-spin system under the fully connected transverse Ising Hamiltonian Eq.~(\ref{Hamiltonian}) is performed in advance to generate teacher data.
    Coefficients \(a_j\) and \(J_{jk}\) are taken from a uniform distribution on \([-1,1]\) independently of the coefficients of Hamiltonian in the circuit.
    The dynamics started from the initialized state \(\ket{0}^{\otimes 10}\).
    The transient at the beginning of evolution is discarded for duration \(T_{\text{transient}} = 300\).
    For practical use, one can employ dynamics obtained experimentally from a quantum system with unknown Hamiltonian as teacher data.
    Learned dynamics is of \(Z\) expectation values of 3 spins during \(t\in [T_{\text{transient}},T_{\text{transient}}+8]\). This span of \(t\) is mapped on \(x\in[-1,1]\) uniformly by \(t = 4(x+1) + T_{\text{transient}}\) to be properly introduced to input gate.
    Output are taken from \(Z\) expectation values of the first, second, and third qubits of the circuit.
    The quadratic cost function is employed.
    The number of teacher samples is 100 for each.
    The multiplication constant \(a\) is fixed to unity.
    
    The result is shown in Fig. \ref{dynamics_fit}.
    It is notable that the 3 observables of a complex 10-spin system can be well fitted, simultaneously, using the 3 observables of a tuned 6-qubit circuit.
    Although the task performed here is not what is commonly referred to as quantum simulation, we believe that we provide an alternative way to learn a quantum many-body dynamics with a near-term quantum computer.
    It may also be possible to extract partial information of the system Hamiltonian by taking derivative of the output with respect to \(x\), which can readily be performed using the same method of calculating a gradient.
   
    \section{Conclusion}
    We have presented a machine learning framework on near-term realizable quantum computers.
    Our method fully employs the exponentially large space of the quantum system, in a way that it mixes simply injected nonlinear functions with a low-depth circuit to approximate a complex nonlinear function.
    Numerical results have shown the ability to represent a function, to classify, and to fit a relatively large quantum system.
    Also, the theoretical investigation has shown QCL's ability to provide us a means to deal with high dimensional regression or classification tasks, which has been unpractical on classical computers.
    We have recently become aware of related works \cite{Cincio2018,Farhi2018,Benedetti2018,Schuld2018,Huggins2018,Liu2018,Schuld2018a,Fanizza2018,Benedetti2018a}.
    
    \appendix*
    \section{Unitarity avoids overfitting}
    In this appendix, we demonstrate a simple example that supports our claim in the main text that states the unitarity of the transformation has an effect to avoid overfittings.
    We perform the one-dimensional fitting task with a small number of training data set to see the avoidance of the overfitting.
    To observe the unitarity effect, we fix the multiplication constant \(a\) to unity.
    For simplicity, here we use a 3-qubit circuit in the same form of the main text, with \(D=3\) and using \(U_{\text{in}} = \prod_i R_i^Y (\sin x)\) as an input gate .
    In this case, the set of basis function that QCL utilizes is \(\{x,x^2,x^3,(1-x^2)^{1/2},1-x^2,(1-x^2)^{3/2},x(1-x^2)^{1/2},x^2(1-x^2)^{1/2},x(1-x^2)\}\).
    Therefore for comparison, we run a simple classical linear regression program using the same basis function set.
    
\begin{figure}[h]
    \includegraphics[width=0.9\linewidth]{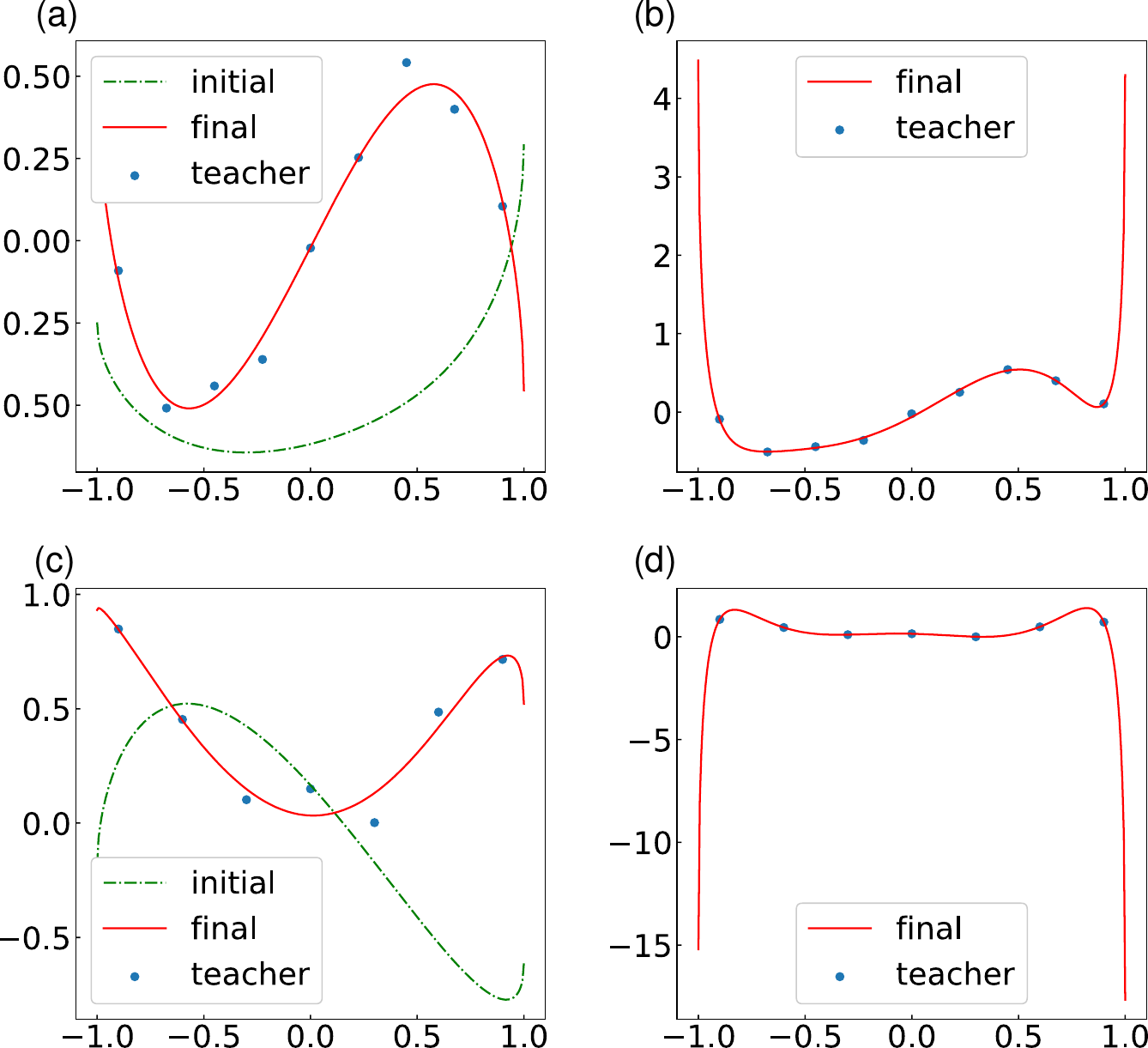}
    \caption{\label{sparse_fit} A simple example of the avoidance of the overfitting resulting from unitarity. (a) and (c): Fitting result of noise-added \(\sin x\) and \(x^2\) using QCL. (b) and (d): Fitting result of noise-added \(\sin x\) and \(x^2\) using the classical regression with same basis functions as used in QCL.}
    \end{figure}

    Fig. \ref{sparse_fit} (a) and (b) show the result of the task to fit data points of \(0.5\sin x\), with Gaussian noise of standard deviation \(0.05\) added, using QCL and classical regression, respectively. The result shows that, probably due to the unitarity of the transformation, QCL accepts some errors in the final output, as opposed to the classical one which does not accept any errors in the final output, that is, it overfits. As opposed to \(\|\bv{w}\|=1\) constraint on QCL, the classical algorithm in this case output a weight vector with \(\|\bv{w}\| \approx 134\). Fig. \ref{sparse_fit} (c) and (d) show the result of the task to fit data points of \(x^2\), with Gaussian noise of standard deviation \(0.05\) added, using QCL and classical regression, respectively. Again, the same observation can be made. The weight vector obtained by the classical algorithm exhibits \(\|\bv{w}\| \approx 15800\) in this case.

   \bibliographystyle{apsrev4-1}

\begin{thebibliography}{36}%
    \makeatletter
    \providecommand \@ifxundefined [1]{%
     \@ifx{#1\undefined}
    }%
    \providecommand \@ifnum [1]{%
     \ifnum #1\expandafter \@firstoftwo
     \else \expandafter \@secondoftwo
     \fi
    }%
    \providecommand \@ifx [1]{%
     \ifx #1\expandafter \@firstoftwo
     \else \expandafter \@secondoftwo
     \fi
    }%
    \providecommand \natexlab [1]{#1}%
    \providecommand \enquote  [1]{``#1''}%
    \providecommand \bibnamefont  [1]{#1}%
    \providecommand \bibfnamefont [1]{#1}%
    \providecommand \citenamefont [1]{#1}%
    \providecommand \href@noop [0]{\@secondoftwo}%
    \providecommand \href [0]{\begingroup \@sanitize@url \@href}%
    \providecommand \@href[1]{\@@startlink{#1}\@@href}%
    \providecommand \@@href[1]{\endgroup#1\@@endlink}%
    \providecommand \@sanitize@url [0]{\catcode `\\12\catcode `\$12\catcode
      `\&12\catcode `\#12\catcode `\^12\catcode `\_12\catcode `\%12\relax}%
    \providecommand \@@startlink[1]{}%
    \providecommand \@@endlink[0]{}%
    \providecommand \url  [0]{\begingroup\@sanitize@url \@url }%
    \providecommand \@url [1]{\endgroup\@href {#1}{\urlprefix }}%
    \providecommand \urlprefix  [0]{URL }%
    \providecommand \Eprint [0]{\href }%
    \providecommand \doibase [0]{http://dx.doi.org/}%
    \providecommand \selectlanguage [0]{\@gobble}%
    \providecommand \bibinfo  [0]{\@secondoftwo}%
    \providecommand \bibfield  [0]{\@secondoftwo}%
    \providecommand \translation [1]{[#1]}%
    \providecommand \BibitemOpen [0]{}%
    \providecommand \bibitemStop [0]{}%
    \providecommand \bibitemNoStop [0]{.\EOS\space}%
    \providecommand \EOS [0]{\spacefactor3000\relax}%
    \providecommand \BibitemShut  [1]{\csname bibitem#1\endcsname}%
    \let\auto@bib@innerbib\@empty
    %</preamble>
    \bibitem [{\citenamefont {Carleo}\ and\ \citenamefont
      {Troyer}(2017)}]{Carleo2017}%
      \BibitemOpen
      \bibfield  {author} {\bibinfo {author} {\bibfnamefont {G.}~\bibnamefont
      {Carleo}}\ and\ \bibinfo {author} {\bibfnamefont {M.}~\bibnamefont
      {Troyer}},\ }\href {\doibase 10.1126/science.aag2302} {\bibfield  {journal}
      {\bibinfo  {journal} {Science}\ }\textbf {\bibinfo {volume} {355}},\ \bibinfo
      {pages} {602} (\bibinfo {year} {2017})}\BibitemShut {NoStop}%
    \bibitem [{\citenamefont {Rupp}\ \emph {et~al.}(2012)\citenamefont {Rupp},
      \citenamefont {Tkatchenko}, \citenamefont {M{\"{u}}ller},\ and\ \citenamefont
      {von Lilienfeld}}]{Rupp2011}%
      \BibitemOpen
      \bibfield  {author} {\bibinfo {author} {\bibfnamefont {M.}~\bibnamefont
      {Rupp}}, \bibinfo {author} {\bibfnamefont {A.}~\bibnamefont {Tkatchenko}},
      \bibinfo {author} {\bibfnamefont {K.-R.}\ \bibnamefont {M{\"{u}}ller}}, \
      and\ \bibinfo {author} {\bibfnamefont {O.~A.}\ \bibnamefont {von
      Lilienfeld}},\ }\href {\doibase 10.1103/PhysRevLett.108.058301} {\bibfield
      {journal} {\bibinfo  {journal} {Phys. Rev. Lett.}\ }\textbf {\bibinfo
      {volume} {108}},\ \bibinfo {pages} {058301} (\bibinfo {year}
      {2012})}\BibitemShut {NoStop}%
    \bibitem [{\citenamefont {Broecker}\ \emph {et~al.}(2017)\citenamefont
      {Broecker}, \citenamefont {Carrasquilla}, \citenamefont {Melko},\ and\
      \citenamefont {Trebst}}]{Broecker2016}%
      \BibitemOpen
      \bibfield  {author} {\bibinfo {author} {\bibfnamefont {P.}~\bibnamefont
      {Broecker}}, \bibinfo {author} {\bibfnamefont {J.}~\bibnamefont
      {Carrasquilla}}, \bibinfo {author} {\bibfnamefont {R.~G.}\ \bibnamefont
      {Melko}}, \ and\ \bibinfo {author} {\bibfnamefont {S.}~\bibnamefont
      {Trebst}},\ }\href {\doibase 10.1038/s41598-017-09098-0} {\bibfield
      {journal} {\bibinfo  {journal} {Sci. Rep.}\ }\textbf {\bibinfo {volume}
      {7}},\ \bibinfo {pages} {8823} (\bibinfo {year} {2017})}\BibitemShut
      {NoStop}%
    \bibitem [{\citenamefont {Ramakrishnan}\ \emph {et~al.}(2015)\citenamefont
      {Ramakrishnan}, \citenamefont {Dral}, \citenamefont {Rupp},\ and\
      \citenamefont {von Lilienfeld}}]{Ramakrishnan2015}%
      \BibitemOpen
      \bibfield  {author} {\bibinfo {author} {\bibfnamefont {R.}~\bibnamefont
      {Ramakrishnan}}, \bibinfo {author} {\bibfnamefont {P.~O.}\ \bibnamefont
      {Dral}}, \bibinfo {author} {\bibfnamefont {M.}~\bibnamefont {Rupp}}, \ and\
      \bibinfo {author} {\bibfnamefont {O.~A.}\ \bibnamefont {von Lilienfeld}},\
      }\href {\doibase 10.1021/acs.jctc.5b00099} {\bibfield  {journal} {\bibinfo
      {journal} {J. Chem. Theory Comput.}\ }\textbf {\bibinfo {volume} {11}},\
      \bibinfo {pages} {2087} (\bibinfo {year} {2015})}\BibitemShut {NoStop}%
    \bibitem [{\citenamefont {August}\ and\ \citenamefont {Ni}(2017)}]{August2016}%
      \BibitemOpen
      \bibfield  {author} {\bibinfo {author} {\bibfnamefont {M.}~\bibnamefont
      {August}}\ and\ \bibinfo {author} {\bibfnamefont {X.}~\bibnamefont {Ni}},\
      }\href {\doibase 10.1103/PhysRevA.95.012335} {\bibfield  {journal} {\bibinfo
      {journal} {Phys. Rev. A}\ }\textbf {\bibinfo {volume} {95}},\ \bibinfo
      {pages} {012335} (\bibinfo {year} {2017})}\BibitemShut {NoStop}%
    \bibitem [{\citenamefont {Shor}(1997)}]{Shor1997}%
      \BibitemOpen
      \bibfield  {author} {\bibinfo {author} {\bibfnamefont {P.~W.}\ \bibnamefont
      {Shor}},\ }\href {\doibase 10.1137/S0097539795293172} {\bibfield  {journal}
      {\bibinfo  {journal} {SIAM J. Comput.}\ }\textbf {\bibinfo {volume} {26}},\
      \bibinfo {pages} {1484} (\bibinfo {year} {1997})}\BibitemShut {NoStop}%
    \bibitem [{\citenamefont {Nielsen}\ and\ \citenamefont
      {Chuang}(2010)}]{Nielsen2010}%
      \BibitemOpen
      \bibfield  {author} {\bibinfo {author} {\bibfnamefont {M.~A.}\ \bibnamefont
      {Nielsen}}\ and\ \bibinfo {author} {\bibfnamefont {I.~L.}\ \bibnamefont
      {Chuang}},\ }\href {\doibase 10.1017/CBO9780511976667} {\emph {\bibinfo
      {title} {{Quantum Computation and Quantum Information}}}}\ (\bibinfo
      {publisher} {Cambridge University Press},\ \bibinfo {address} {Cambridge},\
      \bibinfo {year} {2010})\BibitemShut {NoStop}%
    \bibitem [{\citenamefont {Kerenidis}\ and\ \citenamefont
      {Prakash}()}]{Kerenidis2017}%
      \BibitemOpen
      \bibfield  {author} {\bibinfo {author} {\bibfnamefont {I.}~\bibnamefont
      {Kerenidis}}\ and\ \bibinfo {author} {\bibfnamefont {A.}~\bibnamefont
      {Prakash}},\ }\href {http://arxiv.org/abs/1704.04992} {\ }\Eprint
      {http://arxiv.org/abs/arXiv:1704.04992} {arXiv:1704.04992} \BibitemShut
      {NoStop}%
    \bibitem [{\citenamefont {Wiebe}\ \emph {et~al.}(2012)\citenamefont {Wiebe},
      \citenamefont {Braun},\ and\ \citenamefont {Lloyd}}]{Wiebe2012}%
      \BibitemOpen
      \bibfield  {author} {\bibinfo {author} {\bibfnamefont {N.}~\bibnamefont
      {Wiebe}}, \bibinfo {author} {\bibfnamefont {D.}~\bibnamefont {Braun}}, \ and\
      \bibinfo {author} {\bibfnamefont {S.}~\bibnamefont {Lloyd}},\ }\href
      {\doibase 10.1103/PhysRevLett.109.050505} {\bibfield  {journal} {\bibinfo
      {journal} {Phys. Rev. Lett.}\ }\textbf {\bibinfo {volume} {109}},\ \bibinfo
      {pages} {050505} (\bibinfo {year} {2012})}\BibitemShut {NoStop}%
    \bibitem [{\citenamefont {Rebentrost}\ \emph {et~al.}(2014)\citenamefont
      {Rebentrost}, \citenamefont {Mohseni},\ and\ \citenamefont
      {Lloyd}}]{Rebentrost2014}%
      \BibitemOpen
      \bibfield  {author} {\bibinfo {author} {\bibfnamefont {P.}~\bibnamefont
      {Rebentrost}}, \bibinfo {author} {\bibfnamefont {M.}~\bibnamefont {Mohseni}},
      \ and\ \bibinfo {author} {\bibfnamefont {S.}~\bibnamefont {Lloyd}},\ }\href
      {\doibase 10.1103/PhysRevLett.113.130503} {\bibfield  {journal} {\bibinfo
      {journal} {Phys. Rev. Lett.}\ }\textbf {\bibinfo {volume} {113}},\ \bibinfo
      {pages} {130503} (\bibinfo {year} {2014})}\BibitemShut {NoStop}%
    \bibitem [{\citenamefont {Cao}\ \emph {et~al.}()\citenamefont {Cao},
      \citenamefont {Guerreschi},\ and\ \citenamefont {Aspuru-Guzik}}]{Cao2017}%
      \BibitemOpen
      \bibfield  {author} {\bibinfo {author} {\bibfnamefont {Y.}~\bibnamefont
      {Cao}}, \bibinfo {author} {\bibfnamefont {G.~G.}\ \bibnamefont {Guerreschi}},
      \ and\ \bibinfo {author} {\bibfnamefont {A.}~\bibnamefont {Aspuru-Guzik}},\
      }\href {http://arxiv.org/abs/1711.11240} {\ }\Eprint
      {http://arxiv.org/abs/arXiv:1711.11240} {arXiv:1711.11240} \BibitemShut
      {NoStop}%
    \bibitem [{\citenamefont {Harrow}\ \emph {et~al.}(2009)\citenamefont {Harrow},
      \citenamefont {Hassidim},\ and\ \citenamefont {Lloyd}}]{Harrow2009}%
      \BibitemOpen
      \bibfield  {author} {\bibinfo {author} {\bibfnamefont {A.~W.}\ \bibnamefont
      {Harrow}}, \bibinfo {author} {\bibfnamefont {A.}~\bibnamefont {Hassidim}}, \
      and\ \bibinfo {author} {\bibfnamefont {S.}~\bibnamefont {Lloyd}},\ }\href
      {\doibase 10.1103/PhysRevLett.103.150502} {\bibfield  {journal} {\bibinfo
      {journal} {Phys. Rev. Lett.}\ }\textbf {\bibinfo {volume} {103}},\ \bibinfo
      {pages} {150502} (\bibinfo {year} {2009})}\BibitemShut {NoStop}%
    \bibitem [{\citenamefont {Peruzzo}\ \emph {et~al.}(2014)\citenamefont
      {Peruzzo}, \citenamefont {McClean}, \citenamefont {Shadbolt}, \citenamefont
      {Yung}, \citenamefont {Zhou}, \citenamefont {Love}, \citenamefont
      {Aspuru-Guzik},\ and\ \citenamefont {O'Brien}}]{Peruzzo2013}%
      \BibitemOpen
      \bibfield  {author} {\bibinfo {author} {\bibfnamefont {A.}~\bibnamefont
      {Peruzzo}}, \bibinfo {author} {\bibfnamefont {J.}~\bibnamefont {McClean}},
      \bibinfo {author} {\bibfnamefont {P.}~\bibnamefont {Shadbolt}}, \bibinfo
      {author} {\bibfnamefont {M.}~\bibnamefont {Yung}}, \bibinfo {author}
      {\bibfnamefont {X.}~\bibnamefont {Zhou}}, \bibinfo {author} {\bibfnamefont
      {P.~J.}\ \bibnamefont {Love}}, \bibinfo {author} {\bibfnamefont
      {A.}~\bibnamefont {Aspuru-Guzik}}, \ and\ \bibinfo {author} {\bibfnamefont
      {J.~L.}\ \bibnamefont {O'Brien}},\ }\href
      {http://arxiv.org/abs/1304.3061{\%}0Ahttp://dx.doi.org/10.1038/ncomms5213
      http://www.nature.com/doifinder/10.1038/ncomms5213} {\bibfield  {journal}
      {\bibinfo  {journal} {Nat. Commun.}\ }\textbf {\bibinfo {volume} {5}}
      (\bibinfo {year} {2014})}\BibitemShut {NoStop}%
    \bibitem [{\citenamefont {Kandala}\ \emph {et~al.}(2017)\citenamefont
      {Kandala}, \citenamefont {Mezzacapo}, \citenamefont {Temme}, \citenamefont
      {Takita}, \citenamefont {Brink}, \citenamefont {Chow},\ and\ \citenamefont
      {Gambetta}}]{Kandala2017}%
      \BibitemOpen
      \bibfield  {author} {\bibinfo {author} {\bibfnamefont {A.}~\bibnamefont
      {Kandala}}, \bibinfo {author} {\bibfnamefont {A.}~\bibnamefont {Mezzacapo}},
      \bibinfo {author} {\bibfnamefont {K.}~\bibnamefont {Temme}}, \bibinfo
      {author} {\bibfnamefont {M.}~\bibnamefont {Takita}}, \bibinfo {author}
      {\bibfnamefont {M.}~\bibnamefont {Brink}}, \bibinfo {author} {\bibfnamefont
      {J.~M.}\ \bibnamefont {Chow}}, \ and\ \bibinfo {author} {\bibfnamefont
      {J.~M.}\ \bibnamefont {Gambetta}},\ }\href {\doibase 10.1038/nature23879}
      {\bibfield  {journal} {\bibinfo  {journal} {Nature}\ }\textbf {\bibinfo
      {volume} {549}},\ \bibinfo {pages} {242} (\bibinfo {year}
      {2017})}\BibitemShut {NoStop}%
    \bibitem [{\citenamefont {Farhi}\ \emph {et~al.}()\citenamefont {Farhi},
      \citenamefont {Goldstone},\ and\ \citenamefont {Gutmann}}]{Farhi2014}%
      \BibitemOpen
      \bibfield  {author} {\bibinfo {author} {\bibfnamefont {E.}~\bibnamefont
      {Farhi}}, \bibinfo {author} {\bibfnamefont {J.}~\bibnamefont {Goldstone}}, \
      and\ \bibinfo {author} {\bibfnamefont {S.}~\bibnamefont {Gutmann}},\ }\href
      {http://arxiv.org/abs/1411.4028} {\ }\Eprint
      {http://arxiv.org/abs/arXiv:1411.4028} {arXiv:1411.4028} \BibitemShut
      {NoStop}%
    \bibitem [{\citenamefont {Farhi}\ and\ \citenamefont {Harrow}()}]{Farhi2016}%
      \BibitemOpen
      \bibfield  {author} {\bibinfo {author} {\bibfnamefont {E.}~\bibnamefont
      {Farhi}}\ and\ \bibinfo {author} {\bibfnamefont {A.~W.}\ \bibnamefont
      {Harrow}},\ }\href {http://arxiv.org/abs/1602.07674} {\ }\Eprint
      {http://arxiv.org/abs/arXiv:1602.07674} {arXiv:1602.07674} \BibitemShut
      {NoStop}%
    \bibitem [{\citenamefont {Otterbach}\ \emph {et~al.}(2017)\citenamefont
      {Otterbach}, \citenamefont {Manenti}, \citenamefont {Alidoust}, \citenamefont
      {Bestwick}, \citenamefont {Block}, \citenamefont {Bloom}, \citenamefont
      {Caldwell}, \citenamefont {Didier}, \citenamefont {Fried}, \citenamefont
      {Hong}, \citenamefont {Karalekas}, \citenamefont {Osborn}, \citenamefont
      {Papageorge}, \citenamefont {Peterson}, \citenamefont {Prawiroatmodjo},
      \citenamefont {Rubin}, \citenamefont {Ryan}, \citenamefont {Scarabelli},
      \citenamefont {Scheer}, \citenamefont {Sete}, \citenamefont {Sivarajah},
      \citenamefont {Smith}, \citenamefont {Staley}, \citenamefont {Tezak},
      \citenamefont {Zeng}, \citenamefont {Hudson}, \citenamefont {Johnson},
      \citenamefont {Reagor}, \citenamefont {da~Silva},\ and\ \citenamefont
      {Rigetti}}]{Otterbach2017}%
      \BibitemOpen
      \bibfield  {author} {\bibinfo {author} {\bibfnamefont {J.~S.}\ \bibnamefont
      {Otterbach}}, \bibinfo {author} {\bibfnamefont {R.}~\bibnamefont {Manenti}},
      \bibinfo {author} {\bibfnamefont {N.}~\bibnamefont {Alidoust}}, \bibinfo
      {author} {\bibfnamefont {A.}~\bibnamefont {Bestwick}}, \bibinfo {author}
      {\bibfnamefont {M.}~\bibnamefont {Block}}, \bibinfo {author} {\bibfnamefont
      {B.}~\bibnamefont {Bloom}}, \bibinfo {author} {\bibfnamefont
      {S.}~\bibnamefont {Caldwell}}, \bibinfo {author} {\bibfnamefont
      {N.}~\bibnamefont {Didier}}, \bibinfo {author} {\bibfnamefont {E.~S.}\
      \bibnamefont {Fried}}, \bibinfo {author} {\bibfnamefont {S.}~\bibnamefont
      {Hong}}, \bibinfo {author} {\bibfnamefont {P.}~\bibnamefont {Karalekas}},
      \bibinfo {author} {\bibfnamefont {C.~B.}\ \bibnamefont {Osborn}}, \bibinfo
      {author} {\bibfnamefont {A.}~\bibnamefont {Papageorge}}, \bibinfo {author}
      {\bibfnamefont {E.~C.}\ \bibnamefont {Peterson}}, \bibinfo {author}
      {\bibfnamefont {G.}~\bibnamefont {Prawiroatmodjo}}, \bibinfo {author}
      {\bibfnamefont {N.}~\bibnamefont {Rubin}}, \bibinfo {author} {\bibfnamefont
      {C.~A.}\ \bibnamefont {Ryan}}, \bibinfo {author} {\bibfnamefont
      {D.}~\bibnamefont {Scarabelli}}, \bibinfo {author} {\bibfnamefont
      {M.}~\bibnamefont {Scheer}}, \bibinfo {author} {\bibfnamefont {E.~A.}\
      \bibnamefont {Sete}}, \bibinfo {author} {\bibfnamefont {P.}~\bibnamefont
      {Sivarajah}}, \bibinfo {author} {\bibfnamefont {R.~S.}\ \bibnamefont
      {Smith}}, \bibinfo {author} {\bibfnamefont {A.}~\bibnamefont {Staley}},
      \bibinfo {author} {\bibfnamefont {N.}~\bibnamefont {Tezak}}, \bibinfo
      {author} {\bibfnamefont {W.~J.}\ \bibnamefont {Zeng}}, \bibinfo {author}
      {\bibfnamefont {A.}~\bibnamefont {Hudson}}, \bibinfo {author} {\bibfnamefont
      {B.~R.}\ \bibnamefont {Johnson}}, \bibinfo {author} {\bibfnamefont
      {M.}~\bibnamefont {Reagor}}, \bibinfo {author} {\bibfnamefont {M.~P.}\
      \bibnamefont {da~Silva}}, \ and\ \bibinfo {author} {\bibfnamefont
      {C.}~\bibnamefont {Rigetti}},\ }\href {http://arxiv.org/abs/1712.05771} {\
      (\bibinfo {year} {2017})},\ \Eprint {http://arxiv.org/abs/arXiv:1712.05771}
      {arXiv:1712.05771} \BibitemShut {NoStop}%
    \bibitem [{\citenamefont {Bishop}(2006)}]{Bishop2013}%
      \BibitemOpen
      \bibfield  {author} {\bibinfo {author} {\bibfnamefont {C.~M.}\ \bibnamefont
      {Bishop}},\ }\href
      {http://electronicimaging.spiedigitallibrary.org/article.aspx?doi=10.1117/1.2819119}
      {\emph {\bibinfo {title} {Pattern Recognition and Machine Learning}}}\
      (\bibinfo  {publisher} {Spriger},\ \bibinfo {address} {New York},\ \bibinfo
      {year} {2006})\BibitemShut {NoStop}%
    \bibitem [{\citenamefont {Schuld}\ \emph {et~al.}(2016)\citenamefont {Schuld},
      \citenamefont {Sinayskiy},\ and\ \citenamefont {Petruccione}}]{Schuld2016}%
      \BibitemOpen
      \bibfield  {author} {\bibinfo {author} {\bibfnamefont {M.}~\bibnamefont
      {Schuld}}, \bibinfo {author} {\bibfnamefont {I.}~\bibnamefont {Sinayskiy}}, \
      and\ \bibinfo {author} {\bibfnamefont {F.}~\bibnamefont {Petruccione}},\
      }\href {\doibase 10.1103/PhysRevA.94.022342} {\bibfield  {journal} {\bibinfo
      {journal} {Phys. Rev. A}\ }\textbf {\bibinfo {volume} {94}},\ \bibinfo
      {pages} {022342} (\bibinfo {year} {2016})}\BibitemShut {NoStop}%
    \bibitem [{\citenamefont {Fujii}\ and\ \citenamefont
      {Nakajima}(2017)}]{Fujii2016}%
      \BibitemOpen
      \bibfield  {author} {\bibinfo {author} {\bibfnamefont {K.}~\bibnamefont
      {Fujii}}\ and\ \bibinfo {author} {\bibfnamefont {K.}~\bibnamefont
      {Nakajima}},\ }\href {\doibase 10.1103/PhysRevApplied.8.024030} {\bibfield
      {journal} {\bibinfo  {journal} {Phys. Rev. Appl.}\ }\textbf {\bibinfo
      {volume} {8}},\ \bibinfo {pages} {024030} (\bibinfo {year}
      {2017})}\BibitemShut {NoStop}%
    \bibitem [{\citenamefont {Jaeger}\ and\ \citenamefont
      {Haas}(2004)}]{Jaeger2004}%
      \BibitemOpen
      \bibfield  {author} {\bibinfo {author} {\bibfnamefont {H.}~\bibnamefont
      {Jaeger}}\ and\ \bibinfo {author} {\bibfnamefont {H.}~\bibnamefont {Haas}},\
      }\href {\doibase 10.1126/science.1091277} {\bibfield  {journal} {\bibinfo
      {journal} {Science}\ }\textbf {\bibinfo {volume} {304}},\ \bibinfo {pages}
      {78} (\bibinfo {year} {2004})}\BibitemShut {NoStop}%
    \bibitem [{\citenamefont {Raussendorf}(2005)}]{Raussendorf2004}%
      \BibitemOpen
      \bibfield  {author} {\bibinfo {author} {\bibfnamefont {R.}~\bibnamefont
      {Raussendorf}},\ }\href {\doibase 10.1103/PhysRevA.72.022301} {\bibfield
      {journal} {\bibinfo  {journal} {Phys. Rev. A}\ }\textbf {\bibinfo {volume}
      {72}},\ \bibinfo {pages} {022301} (\bibinfo {year} {2005})},\ \Eprint
      {http://arxiv.org/abs/0412048v1} {0412048v1} \BibitemShut {NoStop}%
    \bibitem [{\citenamefont {Janzing}\ and\ \citenamefont
      {Wocjan}(2005)}]{Janzing2005}%
      \BibitemOpen
      \bibfield  {author} {\bibinfo {author} {\bibfnamefont {D.}~\bibnamefont
      {Janzing}}\ and\ \bibinfo {author} {\bibfnamefont {P.}~\bibnamefont
      {Wocjan}},\ }\href {\doibase 10.1007/s11128-005-4482-9} {\bibfield  {journal}
      {\bibinfo  {journal} {Quantum Inf. Process.}\ }\textbf {\bibinfo {volume}
      {4}},\ \bibinfo {pages} {129} (\bibinfo {year} {2005})}\BibitemShut {NoStop}%
    \bibitem [{\citenamefont {Li}\ \emph {et~al.}(2017)\citenamefont {Li},
      \citenamefont {Yang}, \citenamefont {Peng},\ and\ \citenamefont
      {Sun}}]{Li2017}%
      \BibitemOpen
      \bibfield  {author} {\bibinfo {author} {\bibfnamefont {J.}~\bibnamefont
      {Li}}, \bibinfo {author} {\bibfnamefont {X.}~\bibnamefont {Yang}}, \bibinfo
      {author} {\bibfnamefont {X.}~\bibnamefont {Peng}}, \ and\ \bibinfo {author}
      {\bibfnamefont {C.}~\bibnamefont {Sun}},\ }\href {\doibase
      10.1103/PhysRevLett.118.150503} {\bibfield  {journal} {\bibinfo  {journal}
      {Phys. Rev. Lett.}\ }\textbf {\bibinfo {volume} {118}},\ \bibinfo {pages}
      {150503} (\bibinfo {year} {2017})}\BibitemShut {NoStop}%
    \bibitem [{Note1()}]{Note1}%
      \BibitemOpen
      \bibinfo {note} {The simulation is carried on using Python library QuTip
      \cite {Johansson2013}. We use BFGS method \cite {Nocedal2006} provided in
      SciPy optimization library for optimization of parameters.}\BibitemShut
      {Stop}%
    \bibitem [{\citenamefont {Cincio}\ \emph {et~al.}()\citenamefont {Cincio},
      \citenamefont {Subasi}, \citenamefont {Sornborger},\ and\ \citenamefont
      {Coles}}]{Cincio2018}%
      \BibitemOpen
      \bibfield  {author} {\bibinfo {author} {\bibfnamefont {L.}~\bibnamefont
      {Cincio}}, \bibinfo {author} {\bibfnamefont {Y.}~\bibnamefont {Subasi}},
      \bibinfo {author} {\bibfnamefont {A.~T.}\ \bibnamefont {Sornborger}}, \ and\
      \bibinfo {author} {\bibfnamefont {P.~J.}\ \bibnamefont {Coles}},\ }\href
      {http://arxiv.org/abs/1803.04114} {\ }\Eprint
      {http://arxiv.org/abs/arXiv:1803.04114} {arXiv:1803.04114} \BibitemShut
      {NoStop}%
    \bibitem [{\citenamefont {Farhi}\ and\ \citenamefont {Neven}()}]{Farhi2018}%
      \BibitemOpen
      \bibfield  {author} {\bibinfo {author} {\bibfnamefont {E.}~\bibnamefont
      {Farhi}}\ and\ \bibinfo {author} {\bibfnamefont {H.}~\bibnamefont {Neven}},\
      }\href {http://arxiv.org/abs/1802.06002} {\ }\Eprint
      {http://arxiv.org/abs/arXiv:1802.06002} {arXiv:1802.06002} \BibitemShut
      {NoStop}%
    \bibitem [{\citenamefont {Benedetti}\ \emph {et~al.}()\citenamefont
      {Benedetti}, \citenamefont {Garcia-Pintos}, \citenamefont {Nam},\ and\
      \citenamefont {Perdomo-Ortiz}}]{Benedetti2018}%
      \BibitemOpen
      \bibfield  {author} {\bibinfo {author} {\bibfnamefont {M.}~\bibnamefont
      {Benedetti}}, \bibinfo {author} {\bibfnamefont {D.}~\bibnamefont
      {Garcia-Pintos}}, \bibinfo {author} {\bibfnamefont {Y.}~\bibnamefont {Nam}},
      \ and\ \bibinfo {author} {\bibfnamefont {A.}~\bibnamefont {Perdomo-Ortiz}},\
      }\href {http://arxiv.org/abs/1801.07686} {\ }\Eprint
      {http://arxiv.org/abs/arXiv:1801.07686} {arXiv:1801.07686} \BibitemShut
      {NoStop}%
    \bibitem [{\citenamefont {Schuld}\ and\ \citenamefont
      {Killoran}(2018)}]{Schuld2018}%
      \BibitemOpen
      \bibfield  {author} {\bibinfo {author} {\bibfnamefont {M.}~\bibnamefont
      {Schuld}}\ and\ \bibinfo {author} {\bibfnamefont {N.}~\bibnamefont
      {Killoran}},\ }\href {http://arxiv.org/abs/1803.07128} {\  (\bibinfo {year}
      {2018})},\ \Eprint {http://arxiv.org/abs/arXiv:1803.07128} {arXiv:1803.07128}
      \BibitemShut {NoStop}%
    \bibitem [{\citenamefont {Huggins}\ \emph {et~al.}(2018)\citenamefont
      {Huggins}, \citenamefont {Patel}, \citenamefont {Whaley},\ and\ \citenamefont
      {Stoudenmire}}]{Huggins2018}%
      \BibitemOpen
      \bibfield  {author} {\bibinfo {author} {\bibfnamefont {W.}~\bibnamefont
      {Huggins}}, \bibinfo {author} {\bibfnamefont {P.}~\bibnamefont {Patel}},
      \bibinfo {author} {\bibfnamefont {K.~B.}\ \bibnamefont {Whaley}}, \ and\
      \bibinfo {author} {\bibfnamefont {E.~M.}\ \bibnamefont {Stoudenmire}},\
      }\href {http://arxiv.org/abs/1803.11537} {\  (\bibinfo {year} {2018})},\
      \Eprint {http://arxiv.org/abs/arXiv:1803.11537} {arXiv:1803.11537}
      \BibitemShut {NoStop}%
    \bibitem [{\citenamefont {Liu}\ and\ \citenamefont {Wang}(2018)}]{Liu2018}%
      \BibitemOpen
      \bibfield  {author} {\bibinfo {author} {\bibfnamefont {J.-G.}\ \bibnamefont
      {Liu}}\ and\ \bibinfo {author} {\bibfnamefont {L.}~\bibnamefont {Wang}},\
      }\href {http://arxiv.org/abs/1804.04168} {\  (\bibinfo {year} {2018})},\
      \Eprint {http://arxiv.org/abs/arXiv:1804.04168} {arXiv:1804.04168}
      \BibitemShut {NoStop}%
    \bibitem [{\citenamefont {Schuld}\ \emph {et~al.}(2018)\citenamefont {Schuld},
      \citenamefont {Bocharov}, \citenamefont {Svore},\ and\ \citenamefont
      {Wiebe}}]{Schuld2018a}%
      \BibitemOpen
      \bibfield  {author} {\bibinfo {author} {\bibfnamefont {M.}~\bibnamefont
      {Schuld}}, \bibinfo {author} {\bibfnamefont {A.}~\bibnamefont {Bocharov}},
      \bibinfo {author} {\bibfnamefont {K.}~\bibnamefont {Svore}}, \ and\ \bibinfo
      {author} {\bibfnamefont {N.}~\bibnamefont {Wiebe}},\ }\href
      {http://arxiv.org/abs/1804.00633} {\  (\bibinfo {year} {2018})},\ \Eprint
      {http://arxiv.org/abs/arXiv:1804.00633} {arXiv:1804.00633} \BibitemShut
      {NoStop}%
    \bibitem [{\citenamefont {Fanizza}\ \emph {et~al.}(2018)\citenamefont
      {Fanizza}, \citenamefont {Mari},\ and\ \citenamefont
      {Giovannetti}}]{Fanizza2018}%
      \BibitemOpen
      \bibfield  {author} {\bibinfo {author} {\bibfnamefont {M.}~\bibnamefont
      {Fanizza}}, \bibinfo {author} {\bibfnamefont {A.}~\bibnamefont {Mari}}, \
      and\ \bibinfo {author} {\bibfnamefont {V.}~\bibnamefont {Giovannetti}},\
      }\href {http://arxiv.org/abs/1805.03477} {\  (\bibinfo {year} {2018})},\
      \Eprint {http://arxiv.org/abs/arXiv:1805.03477} {arXiv:1805.03477}
      \BibitemShut {NoStop}%
    \bibitem [{\citenamefont {Benedetti}\ \emph {et~al.}(2018)\citenamefont
      {Benedetti}, \citenamefont {Grant}, \citenamefont {Wossnig},\ and\
      \citenamefont {Severini}}]{Benedetti2018a}%
      \BibitemOpen
      \bibfield  {author} {\bibinfo {author} {\bibfnamefont {M.}~\bibnamefont
      {Benedetti}}, \bibinfo {author} {\bibfnamefont {E.}~\bibnamefont {Grant}},
      \bibinfo {author} {\bibfnamefont {L.}~\bibnamefont {Wossnig}}, \ and\
      \bibinfo {author} {\bibfnamefont {S.}~\bibnamefont {Severini}},\ }\href
      {http://arxiv.org/abs/1806.00463} {\  (\bibinfo {year} {2018})},\ \Eprint
      {http://arxiv.org/abs/arXiv:1806.00463} {arXiv:1806.00463} \BibitemShut
      {NoStop}%
    \bibitem [{\citenamefont {Johansson}\ \emph {et~al.}(2013)\citenamefont
      {Johansson}, \citenamefont {Nation},\ and\ \citenamefont
      {Nori}}]{Johansson2013}%
      \BibitemOpen
      \bibfield  {author} {\bibinfo {author} {\bibfnamefont {J.}~\bibnamefont
      {Johansson}}, \bibinfo {author} {\bibfnamefont {P.}~\bibnamefont {Nation}}, \
      and\ \bibinfo {author} {\bibfnamefont {F.}~\bibnamefont {Nori}},\ }\href
      {\doibase 10.1016/j.cpc.2012.11.019} {\bibfield  {journal} {\bibinfo
      {journal} {Comput. Phys. Commun.}\ }\textbf {\bibinfo {volume} {184}},\
      \bibinfo {pages} {1234} (\bibinfo {year} {2013})}\BibitemShut {NoStop}%
    \bibitem [{\citenamefont {Nocedal}\ and\ \citenamefont
      {Wright}(2006)}]{Nocedal2006}%
      \BibitemOpen
      \bibfield  {author} {\bibinfo {author} {\bibfnamefont {J.}~\bibnamefont
      {Nocedal}}\ and\ \bibinfo {author} {\bibfnamefont {S.}~\bibnamefont
      {Wright}},\ }\href {\doibase 10.1007/978-0-387-40065-5} {\emph {\bibinfo
      {title} {{Numerical Optimization}}}}\ (\bibinfo  {publisher} {Springer},\
      \bibinfo {address} {New York},\ \bibinfo {year} {2006})\BibitemShut {NoStop}%
    \end{thebibliography}

   \end{document}